\documentclass[12]{article}
\usepackage{graphicx,amsmath,amssymb,amsbsy,latexsym, amsfonts, amsthm}
\usepackage[mathscr]{eucal}
\usepackage{bm}

\theoremstyle{plain}
\newtheorem{theorem}{Theorem}
\newtheorem{proposition}[theorem]{Proposition}
\newtheorem{lemma}[theorem]{Lemma}
\newtheorem{corollary}[theorem]{Corollary}
\theoremstyle{definition}
\newtheorem*{definition}{Definition}
\newtheorem*{assumption}{Assumption}
\theoremstyle{remark}

\theoremstyle{plain}
\newcommand{\startproof}{\setlength{\parindent}{0in}\textbf{Proof.} }
\newcommand{\finishproof}{\hfill $\blacksquare$ \\}

\newcommand{\C}{\mathbb{C}}

\newcommand{\R}{\mathbb{R}}

\newcommand{\dif}{\mathrm{d}}

\newcommand{\pderiv}[2]{\frac{\partial #1}{\partial #2}}

\newcommand{\pr}{\mathrm{pr}}
\newcommand{\TDiff}{\mathrm{TDiff}}
\newcommand{\diff}{\mathrm{diff}}
\newcommand{\scrI}{\mathscr{I}}

\newcommand{\Diff}{\mathrm{Diff}}

\newcommand{\lalg}[1]{\mathfrak{#1}}

\newcommand{\Hil}{\mathcal{H}}
\newcommand{\Cyl}{\mathrm{Cyl}}

\newcommand{\Phys}{\mathrm{Phys}}

\newcommand{\Hort}[1]{\mathcal{H}'_{#1}}
\newcommand{\Cylort}[1]{\mathrm{Cyl}'_{#1}}

\newcommand{\dummy}{\rule{0mm}{0mm}}

\newcommand{\scrA}{\mathcal{A}}
\newcommand{\scrD}{\mathcal{D}}
\newcommand{\scrS}{\mathcal{S}}

\begin{document}

\title{Piecewise linear loop quantum gravity}

\author{Jonathan Engle \\
\em Albert-Einstein-Institut, Am M\"{u}hlenberg 1, D-14476 Golm, EU
}
\date{\today}

\maketitle\vspace{-7mm}

\begin{abstract}\noindent

We define a modification of LQG in which graphs are required to consist
in piecewise linear edges, which we call piecewise linear LQG (plLQG).
At the diffeomorphism invariant level, we prove that plLQG is equivalent to
standard LQG, as long as one chooses the class of diffeomorphisms appropriately.
That is, we exhibit a unitary map between the diffeomorphism invariant Hilbert spaces
that maps physically equivalent operators into each other.  In addition, using
the same ideas as in standard LQG, one can define a Hamiltonian and Master constraint in plLQG,
and the unitary map between plLQG and LQG then provides an exact isomorphism of dynamics in
the two frameworks.

Furthermore, loop quantum cosmology (LQC) can be exactly embedded into
plLQG. This allows a prior program of the author to embed LQC
into LQG at the dynamical level to proceed.  In particular, this allows a
formal expression for a physically motivated embedding of LQC into LQG at the
diffeomorphism
invariant level to be given.

\end{abstract}

\section{Introduction}

Loop quantum gravity (LQG) \cite{alrev, otherlqg, lqgorig} is a minimalistic,
background independent approach to quantum gravity.
However, in the construction of the theory, technical choices
have to be made, especially in the kinematics of the theory.
One can then ask: might some of these technical choices
not matter once the constraints are solved?  In this paper,
we show that in particular the choice of the piecewise
analytic category is not essential: it can even be replaced with something
as simple as the piecewise linear category, \textit{and the resulting
theory is the same at the diffeomorphism invariant level}.  The diffeomorphism
invariant Hilbert spaces of the two theories are naturally isomorphic, and the
dynamics are exactly the same. Furthermore, a very large algebra of the
diffeomorphism invariant operators are also seen to be the same.

We call this modification of LQG
`piecewise linear loop quantum gravity' (plLQG).

What are the consequences of this?  First, this can be used
as a ``trick'' to circumvent the obstruction to the program of
\cite{engle2006, engle2007} caused by the non-embeddability
result proved in \cite{bf2007}.  One can arrive at the same
diffeomorphism invariant Hilbert space as in standard LQG,
but by means of a piecewise linear kinematics that completely
circumvents \cite{bf2007}. As a result, it is possible to write
down formal expressions for embeddings of LQC into LQG at the
diffeomorphism invariant level, of the type systematically
motivated in the work \cite{engle2006, engle2007}.
Second, this new framework might allow a closer
relation to spinfoams \cite{othersf, epr},
which also use the piecewise linear
category to define the kinematics \cite{epr}.

After this
work was completed, it was pointed out to the author that the kinematics of piecewise
linear LQG as presented here, and the choice of generalized diffeomorphisms,
had already been proposed as a model in \cite{zapata}.  However \cite{zapata}
was not interested in plLQG as such, and so did not develop it beyond kinematics.
This paper goes further, in rigorously constructing the rigging map
for the diffeomorphism constraint, constructing Hamiltonian and Master constraint operators,
and showing equivalence with the piecewise analytic framework at the diffeomorphism invariant level including
dynamics. Of course the embedding of LQC into plLQG is also new. On the other hand, \cite{zapata}
presents features of the kinematics of plLQG not presented here. For example,
\cite{zapata} introduces the piecewise linear analogue $\overline{\scrA}_{PL}$ of
the generalized connections, and constructs the piecewise linear analogue of the
Ashtekar-Lewandowski measure, allowing one to express the kinematical Hilbert space as
an $L^2$ space.

The paper is organized as follows. First we define the kinematics
of piecewise linear LQG, motivate a choice of generalized diffeomorphism
group, and solve the diffeomorphism constraint.  The unitary map
between the diffeomorphism invariant Hilbert spaces in plLQG and LQG
is then explicitly constructed and proven in section \ref{equivsect}.
Equivalence of diffeomorphism invariant operators in the two frameworks,
and the equivalence of dynamics in the two frameworks is
proven in section \ref{dynamsect}.  The exact embeddings of LQC into plLQG
of the type motivated in \cite{engle2006, engle2007} are then
explicitly reviewed in section \ref{embed_sect}, and at the
end of this section, the resulting formal expressions for
the embeddings of LQC into LQG at the diffeomorphism invariant level are
given. We then close with a brief discussion.

\section{Piecewise linear loop quantum gravity}

\subsection{Kinematics}

We assume space, $M$, is topologically $\R^3$, and we equip
$M$ with a fixed, flat frame bundle connection $\partial_a$.
This flat connection gives us a notion of `straightness' on $M$.
%
%

Let $\scrA$ denote the space of smooth $SU(2)$ connections
on $M$.  The classical phase space is parametrized by such a
connection $A^i_a$ and a densitized triad field $\tilde{E}^a_i$.
(Here $A^i_a$ denotes the components of the $SU(2)$ connection with
respect to the basis $\tau_i:= - \frac{i}{2} \sigma_i$
of the Lie algebra $\lalg{su}(2)$.)
The Poisson brackets are given by
\begin{equation}
\{A^i_a(x), \tilde{E}^b_j(y)\} = 8 \pi \gamma G \delta^i_j \delta^b_a \delta^3(x,y)
\end{equation}
where G is Newton's constant, and
$\gamma \in \R^+$ is the Barbero-Immirzi parameter.

Next one specifies the basic variables.
The algebra of elementary configuration variables is chosen to consist in (real
analytic\footnote{
As always, one has some freedom in the precise definition of cylindrical functions.
This is the definition that will be convenient for this paper.
%
%
}) functions of finite numbers of holonomies of the connection $A^i_a$
along piecewise \textit{straight} edges; we will also use the term
\textit{piecewise linear} for such edges. We call these functions
\textit{piecewise linear cylindrical} and the space of such functions is
denoted $\underline{\Cyl}$.
The elementary momentum variables are
taken to be the \textit{fluxes} on piecewise flat surfaces\footnote{We may
also include the fluxes on arbitrary piecewise analytic surfaces, but nothing
is thereby gained, and using piecewise flat surfaces is more in the spirit of piecewise linear loop quantum gravity as presented here.}.
Given a surface $S$ and a function $f: S \rightarrow \lalg{su}(2)$, we define the corresponding flux by
\begin{equation}
E(S,f) := \int_S f^i \tilde{E}^a_i n_a \dif \sigma_1 \dif \sigma_2
\end{equation}
where $n_a := \epsilon_{abc}\pderiv{x^b}{\sigma_1}\pderiv{x^c}{\sigma_2}$,
$(\sigma_1, \sigma_2)$ are arbitrary coordinates on $S$, $x^a$ are arbitrary
coordinates on the spatial manifold, and $\epsilon_{abc}$ denotes the fully
anti-symmetric symbol (i.e., the Levi-Civita tensor of density weight $-1$).

Next let us introduce some structures to give a more useful
characterization of $\underline{\Cyl}$.
We first define a \textit{piecewise linear path} to
be a continuous path $e: [0,1] \rightarrow M$ consisting in a finite number of
segments, each segment being geodesic with respect to $\partial_a$ (but not necessarily
affinely parametrized.)
%
%
%
We then define a \textit{piecewise linear edge} to be an equivalence class
of piecewise linear paths, where two piecewise linear paths are equivalent
if they are related by a reparametrization, or addition or removal of
`trivial' segment of the form $(\delta \circ \delta^{-1})$.\footnote{Thus,
two paths are equivalent iff they allows yield the same holonomies.}
We next define a \textit{piecewise linear graph} to be a finite, ordered
set of piecewise linear edges.
%
%
Let $\underline{\Gamma}$ denote the space of piecewise linear graphs.
With these definitions, any element $\Phi$ of $\underline{\Cyl}$ can be written
in the form
\begin{equation}
\Phi[A] = F(A(e_1), \dots, A(e_n))
\end{equation}
for some piecewise linear graph $(e_1, \dots, e_n) \in \underline{\Gamma}$,
and some real-analytic
%
%
function $F: SU(2)^n \rightarrow \C$.
If a cylindrical
function $\Phi \in \underline{\Cyl}$ may be written using the edges of a graph
$\gamma$, we say $\Phi$ is \textit{cylindrical with respect to} $\gamma$.
%
%
We denote by $\Cyl_\gamma$ the space of functions cylindrical with respect
to $\gamma$.

We next define an inner product $\langle \cdot, \cdot \rangle$
on $\underline{\Cyl}$ in the same way as in standard LQG:
Given $\Psi, \Phi \in \underline{\Cyl}$, we find a graph $\gamma$
large enough so that $\Psi, \Phi \in \Cyl_\gamma$, and then
define the inner product between $\Psi$ and $\Phi$ using the Haar
measure on $SU(2)$. As in LQG, this inner product is
independent of the ambiguity in the choice of $\gamma$. For each $\gamma$ let
$\Hil_\gamma$ denote the Cauchy completion of $\Cyl_\gamma$,
and let $\underline{\Hil}$ denote the Cauchy completion of $\underline{\Cyl}$,
in this inner product.

We next construct a representation of the basic algebra on $(\underline{\Cyl}, \langle \cdot, \cdot \rangle)$.
The configuration algebra $\underline{\Cyl}$ is represented by multiplication.
The operators corresponding to the momentum degrees of freedom
are then defined via the classical Poisson bracket
\begin{equation}
\widehat{E(S,f)} \Phi = i \{E(S,f), \Phi\}
\end{equation}
which ensures that the commutators of elements of $\underline{\Cyl}$ and
the fluxes match the corresponding Poisson brackets correctly.
The multiplicative $\underline{\Cyl}$ operators are bounded because each element
of $\underline{\Cyl}$, as a continuous function of a finite number of $SU(2)$ holonomies,
is bounded due to the compactness of $SU(2)$. These multiplicative operators
thus extend to all of $\underline{\Hil}$ by the BLT theorem. The
flux operators, equipped with domain $\underline{\Cyl}$, form essentially self-adjoint operators,
which therefore extend uniquely to self-adjoint operators on $\Hil$. One can check that the resulting
representation of the basic observables then reflects correctly not only
the poisson brackets, but also the correct adjointness relations. This is the
elementary quantization.

After the quantization of the elementary operators, other geometrical
operators corresponding to length, area, and volume can also be constructed
in the same way as in standard LQG \cite{alrev, otherlqg, lqggeom}, all with
the same spectra.
The Gauss constraint is defined in the same way as in standard LQG \cite{alrev, otherlqg}
and is just as easy to solve, yielding as a solution space $\underline{\Hil}_{G} \subset \underline{\Hil}$,
consisting in the Cauchy completion of the span of
gauge-invariant spin-network states \cite{alrev, otherlqg}, but this time restricted to graphs in
$\underline{\Gamma}$.

\subsection{Solution to the diffeomorphism contraint}

Next, let us discuss the solution to the diffeomorphism constraint.  Central
to this is the selection of a generalization of the group of diffeomorphisms
to be used in quantum theory.  Once this generalization is selected,
we will simply use the group averaging strategy of \cite{alrev, almmt}
to solve the constraint.

\dummy\\
\noindent\textit{The choice of diffeomorphism gauge group}

Let $\underline{\Diff}$ denote the group of generalized diffeomorphisms to be
used. We first stipulate several requirements of $\underline{\Diff}$, which
will lead us to a choice for the group. First, we stipulate that the
generalized diffeomorphisms at least consist in bijective maps of
space onto itself.\footnote{If one were
to solve the Gauss and diffeomorphism constraints together, this would
be equivalent to requiring that the generalized principal bundle automorphisms to be used
should at least consist in maps from the principal bundle to itself that preserve
all structure of the principal bundle except possibly topology and differentiable structure.}
%
%
Second, each element of $\underline{\Diff}$ must map all piecewise linear
edges to
piecewise linear edges, so that it has a well-defined action on $\underline{\Gamma}$,
the set of piecewise linear graphs.  These requirements, however, so far are not enough:
if we were to only require these, one could map any graph into any other with such a `generalized
diffeomorphism', and, if one follows the prescription of \cite{alrev, almmt}, one would be led to a solution
space with only a single state.
Therefore, we furthermore stipulate that the maps be homeomorphisms.
%
%
A natural choice satisfying the above requirements is the group of
\textit{piecewise linear homeomorphisms}. To define the notion
of a piecewise linear homeomorphism, we must first review the definition of
a simplicial complex \cite{munkres}.  First, we note that the fixed connection $\partial_a$
endows $M$ with a natural \textit{affine structure}.  Let us for
convenience arbitrarily pick an origin $O \in M$, and use this to make $M$
into a vector space, so that addition and real scalar multiplication are defined in $M$.
None of the definitions or constructions below will depend on the choice of $O$.
%
%

A set of points $\{a_0, \dots, a_n\} \subset M$ is said to be \textit{independent}
if they do not lie within any common $(n-1)$-dimensional plane in $M$.  Given
such a set of $n+1$ independent points, we define the \textit{n-simplex} $\sigma$
spanned by $a_0, \dots, a_n$ to be the set of all points $x \in M$ such that
\begin{equation}
x=\sum_{i=0}^n t_i a_i
\end{equation}
for some $t_0, \dots, t_n \in \R$ all non-negative, satisfying $\sum_{i=0}^n t_i = 1$.
$n$ is the called the \textit{dimension} of $\sigma$.
In common language, a 0-simplex is a point, a 1-simplex is a line segment, a 2-simplex
is a triangle, and a 3-simplex is a tetrahedron.

Next we define the generalized notion of `face'.
Given an n-simplex $\sigma$ spanned by a set of points $\{a_0, \dots, a_n\}$, the
simplex spanned by a subset of these points is called a \textit{face} of $\sigma$.
In particular, every simplex is a face of itself; a face of a simplex $\sigma$ that is
not equal to $\sigma$ is called a \textit{proper face}.
Thus, in this generalized sense, the proper `faces' of a tetrahedron consist in all
the triangular faces in the usual sense, all the edges, and all four vertices.
The proper `faces' of a triangle consist in its three edges and three vertices, etc.

We can now define a \textit{simplicial complex} $K$ to be a
(possibly infinite) collection
of simplices such that
\begin{enumerate}
\item Every face of a simplex of $K$ is in $K$.
\item The intersection of any two simplices of K is a face of each of them.
\end{enumerate}
The maximal simplex dimension occuring in $K$ is called the \textit{dimension of $K$}.
%
%
%

Finally, a homeomorphism $F$ from an $n$-dimensional manifold $M$ onto an
$n$-dimensional manifold $N$
is called a \textit{piecewise linear} if there exist simplicial complexes
$K$ and $L$, covering all of $M$ and $N$, respectively, such that $v_1, \dots v_m$ span a simplex
of $K$ if and only if $F(v_1), \dots, F(v_m)$ span a simplex of $L$, and such that
for each $\{v_0, \dots v_n\}$ spanning an $n$-simplex in $K$,
\begin{equation}
F\left(\sum_{i=0}^n t_i v_i\right) = \sum_{i=0}^n t_i F(v_i)
\end{equation}
for all $t_i \ge 0$ satisfying $\sum_{i=0}^n t_i = 1$.
Said simply, $F$ maps simplices of $K$ into simplices of $L$ in a continuous way, such
that $F$ is linear within each n-simplex.\footnote{In
the language of \cite{munkres}, a piecewise linear homeomorphism is
a \textit{simplicial homeomorphism} from some simplicial complex $K$ to another $L$.}

The piecewise linear homeomorphisms are essentially the piecewise linear analogue of the
\textit{stratified analytic diffeomorphisms} advocated in \cite{koslowski2006} and
described in \cite{fleischhack2006, hardt} (see also \cite{fr2004}).
In the analytic framework, however, one has more choices: one
can, for example, require that the generalized diffeomorphisms be at least differentiable.
The analogue of such a requirement can, however, \textit{not}
be satified in the piecewise linear framework: the only differentiable piecewise linear
maps are fully linear.  But the group of globally linear maps is too small:
if one were to choose $\underline{\Diff}$ to be the group of linear maps, even individual
open edges would have global information that would be diffeomorphism-invariant.
This would prevent any possible relation, with any analytic LQG framework so far proposed,
at the diffeomorphism-invariant level.
%
%

\dummy\\
\textit{Construction of the diffeomorphism invariant Hilbert space}

With the foregoing choice of $\underline{\Diff}$, let us proceed to construct the
solution to the diffeomorphism constraint.  For this purpose, we introduce some further
definitions.
First, if two graphs $\gamma_1, \gamma_2 \in \underline{\Gamma}$ differ only by
a permutation of edges or reversal of edge orientations, call them \textit{probe equivalent}.
The probe equivalence class of a graph $\gamma$ we write as $[\gamma]_{pr}$.
Let $\underline{\Gamma}_\pr$ denote the space of such probe equivalence classes in $\underline{\Gamma}$.
Next, for each $\gamma \in \underline{\Gamma}$, let $\Hort{\gamma}$ denote the
orthogonal complement, in $\Hil_\gamma$, of the span of all functions that are constant
on at least one edge of $\gamma$. Then, as in \cite{alrev},
\begin{equation}
\underline{\Hil} = \oplus_{[\gamma]_{\pr} \in \underline{\Gamma}_\pr} \Hort{\gamma}.
\end{equation}
Furthermore let $\Cylort{\gamma}:=\Hort{\gamma} \cap \Cyl$.
Lastly we define some subgroups of our chosen generalized diffeomorphisms.
For each $\gamma \in \underline{\Gamma}$, let $\underline{\Diff}_\gamma$ be the set of elements
in $\underline{\Diff}$ mapping $\gamma$ back into its probe equivalence class.
Let $\underline{\TDiff}_\gamma$ be the set of elements in $\underline{\Diff}$ fixing $\gamma$, so that
they preserve each edge of $\gamma$ including orientation.
Let $\underline{GS}_\gamma:= \underline{\Diff}_\gamma/ \underline{\TDiff}_\gamma$ where
the division is taken with respect to the left-action.

For each $\gamma \in \underline{\Gamma}$, define $\underline{P}_{\diff,\gamma}$ as the
\textit{group averaging map} \cite{almmt, alrev} from $\Hort{\gamma}$ to the subspace invariant under $\underline{GS}_\gamma$:\footnote{In lemma \ref{avemaps_eq}, we will show
$\underline{P}_{\diff,\gamma}$ is equal to $P_{\diff,\gamma}$ in \cite{alrev}.}
\begin{equation}
\underline{P}_{\diff,\gamma} \Psi_\gamma := \frac{1}{\left|\underline{GS}_\gamma \right|}
\sum_{\varphi \in \underline{GS}_\gamma} \varphi^* \Psi_\gamma .
\end{equation}
For each $\Psi_\gamma \in \Cylort{\gamma}$, define $\underline{\eta}(\Psi_\gamma) \in \underline{\Cyl}^*$ by
\begin{equation}
\label{pl_eta}
(\underline{\eta}(\Psi_\gamma)|\Phi \rangle
:= \sum_{\varphi \in \underline{\Diff}/\underline{\Diff}_\gamma \hspace{-3em}\dummy} \langle \varphi^* \underline{P}_{\diff, \gamma} \Psi_\gamma, \Phi \rangle
= \frac{1}{\left|\underline{GS}_\gamma \right|} \sum_{\varphi \in \underline{\Diff}/\underline{\TDiff}_\gamma \hspace{-3em}\dummy} \langle
\varphi^* \Psi_\gamma, \Phi \rangle .
\end{equation}
Piecing together these maps for the various $\gamma \in \underline{\Gamma}$ defines a map $\underline{\eta}: \underline{\Cyl} \rightarrow \underline{\Cyl}^*$.
This is the rigging map for solving the diffeomorphism constraint for piecewise linear LQG.
The space of `test functions' at the diffeomorphism invariant level is then
\begin{equation}
\underline{\Cyl}_{\diff}^*:= {\rm Im} \underline{\eta}.
\end{equation}
The inner product on this space is defined as follows:
For $\underline{\eta} \Psi, \underline{\eta} \Phi \in {\rm Im} \underline{\eta}$,
\begin{equation}
      \langle \underline{\eta} \Psi, \underline{\eta} \Phi \rangle := ( \underline{\eta} \Psi | \Phi \rangle.
\end{equation}
The Cauchy completion of $\underline{\Cyl}_{\diff}^*$ with respect to the above inner product we denote by $\underline{\Hil}_\diff$.

The solution to both the Gauss and diffeomorphism constraints is constructed by first defining
$\underline{\Cyl}_{\diff, G}^* := \underline{\eta}[\underline{\Cyl} \cap \underline{\Hil}_G] \subset \underline{\Cyl}_\diff^*$,
and then Cauchy completing to obtain $\underline{\Hil}_{\diff, G} \subset \underline{\Hil}_\diff$.

\section{Equivalence of piecewise linear LQG with analytic LQG
at the diffeomorphism invariant level}
\label{equivsect}

In this section we prove that the diffeomorphism invariant Hilbert space
for piecewise linear LQG is naturally isomorphic to the
diffeomorphism invariant Hilbert space of standard LQG --- provided that
for standard LQG one uses a generalized diffeomorphism group such as that
advocated by \cite{koslowski2006}.

We begin by proving the key lemma about piecewise linear LQG allowing the
equivalence.  Essentially it states that $\underline{\Diff}$ equivalence classes
of piecewise linear graphs are simply knot classes.  Because the analogue of this
is also true for piecewise analytic LQG with the choice of diffeomorphism
group advocated in \cite{koslowski2006}, one already has a hint of the equivalence
of the two theories at the diffeomorphism invariant level.  However, to rigorously
prove the equivalence, more must be done, and the subsequent part of this section is
devoted to this task.

First we give several definitions.
Given a simplicial complex $K$,
a \textit{subcomplex} $K'$ is any subset of $K$ such
that $K'$ is again a simplicial complex. (Note it is possible
for the dimension of $K'$ to be less than that of $K$).  Second,
a complex $\tilde{K}$ is said to be a \textit{subdivision} of a complex $K$
if every simplex of $\tilde{K}$ is contained in a simplex of $K$, and every simplex of $K$
is a union of simplices in $\tilde{K}$.  Third, given a simplicial complex $K$,
we define
\begin{equation}
|K|:= \cup_{A \in K} A,
\end{equation}
called the \textit{polyhedron underlying} $K$.
Lastly, we define a piecewise linear graph $\gamma$ and a 1-complex $X$ to be \textit{compatible} if the image of $\gamma$ (which we denote by $|\gamma|$) equals $|X|$.
By breaking up each edge of a piecewise linear $\gamma$ into its straight
pieces, and taking the set of these line segments and all their endpoints,
one obtains the simplest 1-complex compatible with $\gamma$.
%
%
%
By
subdividing the edges further, one obtains other compatible 1-complexes.
%
%
%
%

We begin by stating a lemma, which is almost identical to (4.4) of \cite{brown}:
\begin{lemma}[almost (4.4) of Brown \cite{brown}]
\label{from_br44}
Let $K$ and $L$ be 3-complexes and let $K_1$ and $L_1$ be 1-dimensional subcomplexes
of $K$ and $L$ respectively.  Suppose $f: |K| \rightarrow |L|$ is a homeomorphism such
that $f(|K_1|)=|L_1|$. Then there exists an isotopy $g_t:|K| \rightarrow |L|$ such that
%
%
%
\begin{enumerate}
\item[(i)] $g_0 = f$

\item[(ii)] there exist subdivisions $\tilde{K}, \tilde{L}, \tilde{K}_1, \tilde{L}_1$ of
$K, L, K_1, L_1$ respectively such that
\begin{enumerate}
\item[(a.)] $\tilde{K}_1$ and $\tilde{L}_1$ are subcomplexes of $\tilde{K}$ and $\tilde{L}$, respectively,
\item[(b.)] $g_t$ maps $\tilde{K}_1$ onto $\tilde{L}_1$ for all $t$, and

\item[(c.)] $g_1$ is piecewise linear on $\tilde{K}_1$.
\end{enumerate}
\end{enumerate}
\end{lemma}
{\startproof
The proof is exactly the same as that given for (4.4) in \cite{brown};
only the statement of the lemma differs.
\finishproof}
We use the above in proving the following lemma.
A generalized version of the Hauptvermutung of algebraic topology
for 3-complexes, proved in 1969 \cite{brown}, plays a key role in
the following proof.
\begin{lemma}
\label{plgraphtrans}
If $\gamma, \gamma' \in \underline{\Gamma}$ admit a homeomorphism $\xi:M \rightarrow M$ such that
$\gamma' = \xi \cdot \gamma$, then there exists $\varphi \in \underline{\Diff}$
such that $\gamma' = \varphi \cdot \gamma$.
\end{lemma}
{\startproof

First, by theorem \ref{triangexists} in the appendix, there exist simplicial complexes
$K$ and $L$, each triangulating all of $M = \R^3$, such that
$K$ contains a one-dimensional subcomplex $K_1$ compatible with $\gamma$,
and $L$ contains a one-dimensional subcomplex $L_1$ compatible with $\gamma'$.
Because $\xi$ maps $\gamma$ to $\gamma'$, it maps $|K_1|$ to $|L_1|$.
We now invoke lemma \ref{from_br44} above; it provides us with subdivisions
$\tilde{K}, \tilde{L}, \tilde{K}_1, \tilde{L}_1$ of $K, L, K_1, L_1$
such that $\tilde{K}_1$ and $\tilde{L}_1$ are subcomplexes of $\tilde{K}$ and $\tilde{L}$,
and an isotopy $\xi_t:M \rightarrow M$ such that (i) $\xi_0 = \xi$, (ii) $\xi_t$ maps $\tilde{K}_1$ to $\tilde{L}_1$
for all $t$, and (iii) $\xi_1$ is piecewise linear on $\tilde{K}$.

The 3-complexes $\tilde{K}$ and $\tilde{L}$, the subcomplex $\tilde{K}_1$ of $\tilde{K}$,
and the homeomorphism $\xi_1$ now satisfy the hypotheses
of theorem (4.8) of \cite{brown}, which implies the existence of 
an isotopy $\varphi_t : M \rightarrow M$, such that (i) $\varphi_0 = \xi_1$,
(ii) $\varphi_1$ is piecewise linear, and (iii) $\varphi_t|_{|\tilde{K}_1|} = \xi_1|_{|\tilde{K}_1|}$
for all $t$.

Now, as already noted, $\xi_t$ maps $\tilde{K}_1$ as a 1-complex onto $\tilde{L}_1$
as a 1-complex for all $t$. That is, $\xi_t$ maps each simplex of $\tilde{K}_1$ to a corresponding
simplex of $\tilde{L}_1$ in an onto fashion; this mapping is furthermore 1-1 from the injectivity of $\xi_t$.
Now, because $\tilde{K}_1$ is a subdivision of $K_1$, and $K_1$ is compatible with $\gamma$,
$\tilde{K}_1$ is also compatible with $\gamma$, so that each edge of
$\gamma$ is a union if simplices in $\tilde{K}_1$.  Likewise, each edge of $\gamma'$ is a union
of simplices in $\tilde{L}_1$.  It follows that, for all $t$, $\xi_t$ maps each edge of $\gamma$
onto a corresponding edge of $\gamma'$ in a 1-1 and onto fashion.  The continuity of $\xi_t$ in $t$
ensures that $\xi_t$ always maps each edge of $\gamma$ to the \textit{same} edge of $\gamma'$ for
all $t$.  Furthermore, recall that $\xi_0 = \xi$ maps the orientation of each edge in $\gamma$ correctly
into the orientation of the corresponding edge in $\gamma'$; the continuity of $\xi_t$ in $t$
ensures that $\xi_t$ does the same for all $t$.  Thus, for all $t$, $\xi_t$ maps $\gamma$ onto
$\gamma'$ as a graph.  This is in particular true for $\xi_1$; property (iii) of $\varphi_t$ then
implies that this is also true for $\varphi_t$ for all $t$. $\varphi:=\varphi_1$ thus provides a piecewise linear homeomorphism, i.e., an element of $\underline{\Diff}$, mapping $\gamma$ to $\gamma'$, as desired.
\finishproof}

Let $\Gamma$ denote the set of piecewise analytic graphs: that is, graphs with a finite number
of oriented compact edges, each of which can be subdivided into a finite number of analytic curves.
\begin{definition}[probe equivalent]
When two graphs $\gamma, \gamma' \in \Gamma$ differ only by
a permutation of edges or reversal of edge orientations, we say that $\gamma$ and $\gamma'$
are \textit{probe equivalent}.  The probe equivalence class of a graph $\gamma$ we write $[\gamma]_{pr}$.
\end{definition}
\noindent Let $\Gamma_\pr$ denote the set of probe equivalence classes in $\Gamma$,
as we have let $\underline{\Gamma}_\pr$ denote the set of probe equivalence classes in $\underline{\Gamma}$.
%
%
Let $\Diff$ denote the class of diffeomorphisms which one wishes to use
to solve the diffeomorphism constraint in the piecewise analytic framework.  We make the following
assumption about $\Diff$:
\begin{assumption}
\label{angraphtrans}
If $\gamma, \gamma' \in \Gamma$ are such that $\gamma' = \xi \cdot \gamma$
for some homeomorphism $\xi: M \rightarrow M$, then there exists $\varphi \in \Diff$
such that $\gamma' = \varphi \cdot \gamma$.
\end{assumption}
\noindent Note that if $\Diff$ is chosen to be the stratified analytic diffeomorphisms \cite{fleischhack2006, hardt}
as advocated in \cite{koslowski2006}, lemma 4 in \cite{koslowski2006}\footnote{using the
analytic differentiability class} \textit{assures that this assumption is satisfied}.
Finally, let $\scrA$ denote the space of smooth $SU(2)$ connections on $M$.
In defining analytic LQG and its diffeomorphism-invariant Hilbert-space, we follow
\cite{alrev}.  In the following, we only introduce the structures necessary
to construct the diffeomorphism invariant Hilbert space.
%
%
\begin{definition}[Piecewise analytic LQG structures]
\dummy
\begin{enumerate}

\item Given a graph $\gamma \in \Gamma$, let $\Cyl_\gamma$ denote the set of functions on $\scrA$
cylindrical with respect to $\gamma$ (note that for $\gamma \in \underline{\Gamma}$, this is
consistent with the prior definition of $\Cyl_\gamma$). Let $\Cyl:= \cup_{\gamma} \Cyl_\gamma$

\item Let $\langle , \rangle$ denote the standard inner product on $\Cyl$ defined
using the Haar measure on $SU(2)$ \cite{alrev, otherlqg}.  Let $\Hil_\gamma$
and $\Hil$ denote the Cauchy completions
of $\Cyl_\gamma$ and $\Cyl$, respectively, with respect to $\langle, \rangle$.

\item Let $\Hil_G$ denote the solution space to the Gauss constraint, consisting as usual
in the Cauchy completion of the span of gauge-invariant spin-networks\cite{alrev, otherlqg}.

\item For each $\gamma \in \Gamma$, let $\Hort{\gamma}$ denote the orthogonal complement, in
$\Hil_\gamma$, of the span of all functions that are constant on at least one edge
of $\gamma$, so that, as in \cite{alrev},
$\Hil = \oplus_{[\gamma] \in \Gamma_\pr} \Hort{\gamma}$.  Let $\Cylort{\gamma}:= \Cyl \cap \Hort{\gamma}$.
(For $\gamma \in \overline{\Gamma}$, these definitions are again consistent with the ones
in the piecewise linear framework.)

\item For each $\gamma \in \Gamma$, let $\Diff_\gamma$ be the set of elements
in $\Diff$ mapping $\gamma$ back into its probe equivalence class.
Let $\TDiff_\gamma$ be the set of elements in $\Diff$ that fix $\gamma$ ---
i.e., that preserve each edge of $\gamma$, including orientation.
So defined, $\Diff_\gamma$ is precisely the subset of $\Diff$ preserving
$\Cylort{\gamma}$ under pull-back, and $\TDiff_\gamma$ is precisely the subset of
$\Diff$ that acts as the identity on $\Cylort{\gamma}$ under pull-back.
Let $GS_\gamma:= \Diff_\gamma/ \TDiff_\gamma$ where
the division is taken with respect to the left-action.

\item For each $\gamma \in \Gamma$, define $P_{\diff,\gamma}$ as the
group averaging map from $\Hort{\gamma}$ to the subspace invariant under $GS_\gamma$:
\begin{equation}
P_{\diff,\gamma} \Psi_\gamma := \frac{1}{\left|GS_\gamma \right|} \sum_{\varphi \in GS_\gamma} \varphi^* \Psi_\gamma .
\end{equation}
For each $\Psi_\gamma \in \Cylort{\gamma}$, define $\eta(\Psi_\gamma) \in \Cyl^*$ by
\begin{equation}
\label{an_eta}
(\eta(\Psi_\gamma)|\Phi \rangle
:= \sum_{\varphi \in \Diff/\Diff_\gamma \hspace{-3em}\dummy} \langle \varphi^* P_{\diff, \gamma} \Psi_\gamma, \Phi \rangle
= \frac{1}{\left|GS_\gamma \right|}\sum_{\varphi \in \Diff/\TDiff_\gamma \hspace{-3em}\dummy} \langle \varphi^* \Psi_\gamma, \Phi \rangle .
\end{equation}
Piecing these together for all $\gamma$ defines a map $\eta: \Cyl \rightarrow \Cyl^*$.
This is the rigging map for the theory, as defined in \cite{alrev} (see also \cite{almmt}, and the related
\cite{fr2004}).

\item $\Cyl_{\diff}^*:= {\rm Im} \eta$.  For $\eta \Psi, \eta \Phi \in {\rm Im} \eta$,
    \begin{equation}
      \langle \eta \Psi, \eta \Phi \rangle := ( \eta \Psi | \Phi \rangle .
    \end{equation}
    $\Hil_\diff$ is then defined to be the Cauchy completion of $\Cyl_{\diff}^*$ with respect to
    this inner product.  The completion $\Hil_{\diff, G}$ of the subspace
    $\Cyl_{\diff, G}^* := \eta[\Cyl \cap \Hil_G] \subset \Hil_\diff$ is then
    the solution to both the Gauss and diffeomorphism constraints.

\end{enumerate}
\end{definition}
We next prove a few important lemmas which we use.
\begin{lemma}
\label{Fiso}
For each $\gamma \in \underline{\Gamma}$, the map
\begin{eqnarray}
\nonumber
F: &\underline{GS}_\gamma &\rightarrow GS_\gamma \\
   &\varphi \circ \left[\underline{\TDiff}_\gamma\right] &\mapsto \varphi \circ \left[\TDiff_\gamma\right]
\end{eqnarray}
is well-defined, and is an isomorphism, showing $\underline{GS}_\gamma \cong GS_\gamma$.
\end{lemma}
{\startproof

\textit{F is well-defined}\\
Suppose $\varphi, \xi \in \underline{\Diff}_\gamma$ are such that
$\varphi \circ \left[\underline{\TDiff}_\gamma\right] = \xi \circ \left[\underline{\TDiff}_\gamma\right]$.
Then $\varphi^{-1} \circ \xi \in \underline{\TDiff}_\gamma$, whence $\varphi^{-1} \circ \xi \in \TDiff_\gamma$ also, so that
$\varphi \circ \left[\TDiff_\gamma\right] = \xi \circ \left[\TDiff_\gamma\right]$, proving $F$ well-defined.\\
\textit{F is a homomorphism}\\
This is immediate from the definition of multiplication in the two quotient groups.\\
\textit{F is injective}\\
Suppose $\varphi, \xi \in \underline{\Diff}$ are such that
$\varphi \circ \left[\TDiff_\gamma\right] = \xi \circ \left[\TDiff_\gamma\right]$.
Then $\varphi^{-1} \circ \xi \in \TDiff_\gamma$.  But $\varphi, \xi \in \underline{\Diff}$, so that
$\varphi^{-1} \circ \xi \in \underline{\Diff}$, proving furthermore $\varphi^{-1} \circ \xi \in \underline{\TDiff}_\gamma$.
It follows $\varphi \circ \left[\underline{\TDiff}_\gamma\right] = \xi \circ \left[\underline{\TDiff}_\gamma\right]$, proving
injectivity.\\
\textit{F is surjective}\\
Let $\varphi \circ \left[\TDiff_\gamma\right] \in GS_\gamma$ be given, so that $\varphi \in \Diff_\gamma$.
Let $\gamma' := \varphi \cdot \gamma$.  As $\varphi \in \Diff_\gamma$, $\gamma'$ is probe equivalent to
$\gamma$ and so is also in $\underline{\Gamma}$.
%
%
Furthermore, $\varphi$ is in particular a homeomorphism, allowing us to invoke
lemma \ref{plgraphtrans}, so that there exists a $\xi \in \underline{\Diff}$ such that
$\gamma' = \xi \cdot \gamma$.  This $\xi$ maps $\gamma$ to $\gamma'$, a graph probe equivalent
to $\gamma$, whence $\xi \in \underline{\Diff}_\gamma$.  Furthermore,
$(\varphi^{-1} \circ \xi) \gamma = \gamma$, so that $\varphi^{-1} \circ \xi \in \TDiff_\gamma$,
whence $\xi \circ [\TDiff_\gamma] = \varphi \circ [\TDiff_\gamma]$.  Thus
$F(\xi \circ [\underline{\TDiff}_\gamma]) = \varphi \circ [\TDiff_\gamma]$,
proving surjectivity.
\finishproof}

\begin{lemma}
\label{cylorbit}
\dummy
\begin{enumerate}
\item
Given $\gamma \in \Gamma$, there exists $\varphi \in \Diff$ such that $\varphi \cdot \gamma \in \underline{\Gamma}$.
\item
Given $\Psi \in \Cyl$, there exists $\tilde{\varphi} \in \Diff$
such that $\tilde{\varphi}^* \Psi \in \underline{\Cyl}$.
\end{enumerate}
\end{lemma}
{\startproof

\textit{Proof of (1.)}:\\
Let $\alpha$ be any element of $\underline{\Gamma}$ with the same knot-class
as $\gamma$ (it easy to see that one can construct an element of $\underline{\Gamma}$
with any desired knot-class), and choose the ordering and orientation of the edges
of $\alpha$ such that $\alpha = \xi \cdot \gamma$ for some homeomorphism $\xi: M \rightarrow M$.
Assumption \ref{angraphtrans} implies there exists $\varphi \in \Diff$ such that $\alpha = \varphi \cdot \gamma$.

\textit{Proof of (2.)}:\\
As $\Psi \in \Cyl$, $\Psi \in \Cyl_\gamma$ for some $\gamma \in \Gamma$.
From part (1.), there exists $\varphi \in \Diff$ such that $\varphi \cdot \gamma \in \underline{\Gamma}$, so that $(\varphi^{-1})^* \Psi \in \underline{\Cyl}$.

\finishproof}
Because $\underline{\Cyl} \subset \Cyl$, we have a natural map $\scrI: \Cyl^* \rightarrow \underline{\Cyl}^*$
defined by
\begin{equation}
(\scrI \Psi | \Phi \rangle := (\Psi | \Phi \rangle
\end{equation}
for all $\Psi \in \Cyl^*$ and $\Phi \in \underline{\Cyl}$.

\begin{lemma}
\label{avemaps_eq}\dummy
\begin{enumerate}
\item
\label{P_eq}
For $\gamma \in \underline{\Gamma}$, $P_{\diff, \gamma} = \underline{P}_{\diff, \gamma}$.

\item
\label{eta_eq}
For $\Psi \in \underline{\Cyl}$, $\scrI \eta \Psi = \underline{\eta} \Psi$.
\end{enumerate}
\end{lemma}
{\startproof

\textit{Proof of (\ref{P_eq}.):}\\
We use the isomorphism $F$ from lemma \ref{Fiso}.  It is immediate from its definition that,
for $\Psi \in \Cylort{\gamma}$ and
$\xi \in \underline{\Diff}_\gamma/\underline{\TDiff}_\gamma$,
$F(\xi)^* \Psi = \xi^* \Psi$.  Using $F$ and this fact,
\begin{eqnarray}
P_{\diff, \gamma} \Psi := \frac{1}{\left|GS_\gamma\right|} \sum_{\varphi \in GS_\gamma} \varphi^* \Psi
= \frac{1}{\left|\underline{GS}_\gamma\right|} \sum_{\varphi \in \underline{GS}_\gamma} (F\varphi)^* \Psi
= \frac{1}{\left|\underline{GS}_\gamma\right|} \sum_{\varphi \in \underline{GS}_\gamma} \varphi^* \Psi
= \underline{P}_{\diff, \gamma} \Psi .
\end{eqnarray}

\textit{Proof of (\ref{eta_eq}):}\\
Using the linearity of $\scrI$, $\eta$ and $\eta'$, without loss of generality, assume
$\Psi \in \Cylort{\gamma}$ for some $\gamma \in \underline{\Gamma}$.
Suppose $\gamma' \in \underline{\Gamma}$ and $\Theta \in \Cylort{\gamma'}$ are given.\\
\textit{Case 1:} There exists no $\varphi_o \in \Diff$ such that $\varphi_o \cdot \gamma' = \gamma$.
\begin{quote}
Then from (\ref{an_eta}), $(\scrI \eta \Psi \mid \Theta\rangle = 0$.  But from (\ref{pl_eta}),
$(\underline{\eta} \Psi \mid \Theta \rangle = 0$ as well, so that
$(\scrI \eta \Psi \mid \Theta \rangle = (\underline{\eta} \Psi \mid \Theta \rangle = 0$.
\end{quote}
\textit{Case 2:} There exists $\varphi_o \in \Diff$ such that $\varphi_o \cdot \gamma' = \gamma$.
\begin{quote}
Then, from lemma \ref{plgraphtrans}, there exists $\underline{\varphi}_o \in \underline{\Diff}$
such that $\underline{\varphi}_o \cdot \gamma' = \gamma$.
Using the orthogonality of the spaces $\Hort{\gamma}$,
the middle expression in (\ref{an_eta}) reduces to
\begin{equation}
(\scrI \eta \Psi | \Theta \rangle = (\eta \Psi | \Theta \rangle
= \langle \underline{\varphi}_o^* P_{\diff, \gamma}\Psi, \Theta \rangle
\end{equation}
Using part (\ref{P_eq}.) of this lemma, and the same orthogonality of the spaces $\Hort{\gamma}$
to simplify the expression for $(\underline{\eta}\Psi \mid \Theta\rangle$, we also have
\begin{equation}
(\scrI \eta \Psi | \Theta \rangle
= \langle \underline{\varphi}_o^* \underline{P}_{\diff, \gamma}\Psi, \Theta \rangle
= (\underline{\eta} \Psi | \Theta \rangle
\end{equation}
\end{quote}
Thus $(\scrI \eta \Psi \mid \Theta \rangle = (\underline{\eta} \Psi \mid \Theta \rangle$
for all $\Theta \in \Cylort{\gamma'}$, $\gamma' \in \underline{\Gamma}$, so that
$\scrI \eta \Psi = \underline{\eta} \Psi$.
\finishproof}

\begin{theorem}
\label{diffiso}
$\scrI$ maps $\Cyl_{\diff}^*$ onto $\underline{\Cyl}_{\diff}^*$.
Furthermore,
$\scrI |_{\Cyl_{\diff}^*}: \Cyl_{\diff}^* \rightarrow \underline{\Cyl}_{\diff}^*$ is a unitary isomorphism.
\end{theorem}
{\startproof

\textit{Proof that $\scrI[\Cyl_{\diff}^*]=\underline{\Cyl}_{\diff}^*$}:
\begin{quote}
\textit{$\subseteq$}:\\
Let $\eta \Psi \in \Cyl_{\diff}^*$ be given, so that
$\Psi \in \Cyl$. By lemma \ref{cylorbit}, $\exists \xi \in \Diff$
s.t. $\xi^* \Psi \in \underline{\Cyl}$. Using the $\Diff$
invariance of $\eta$ and part (2.) of lemma \ref{avemaps_eq},
$\scrI \eta \Psi = \scrI \eta (\xi^* \Psi) =
\underline{\eta}(\xi^* \Psi)$, which is in $\underline{\Cyl}_{\diff}^*$.\\
\textit{$\supseteq$}:\\
Let $\underline{\eta} \Psi \in \underline{\Cyl}_{\diff}^*$ be given, so $\Psi \in \underline{\Cyl}$.
Then $\eta \Psi \in \Cyl_{\diff}^*$, and by lemma \ref{avemaps_eq},
$\scrI \eta \Psi = \underline{\eta} \Psi$, so that
$\underline{\eta} \Psi \in \scrI\left[\Cyl_{\diff}^*\right]$.
\end{quote}

%
\textit{Proof that $\scrI|_{\Cyl_\diff^*}$ is injective:}
\begin{quote}
Suppose $\eta \Psi, \eta \Phi \in \Cyl_{\diff}^*$ are such
that $\scrI \eta \Psi = \scrI \eta \Phi$.
Let $\Theta \in \Cyl$ be given. By lemma \ref{cylorbit},
there exists $\xi \in \Diff$ such that $\xi^* \Theta \in \underline{\Cyl}$.
Using the $\Diff$ invariance of $\eta$,
%
%
\begin{displaymath}
(\eta \Psi | \Theta \rangle
= (\eta \Psi | \xi^* \Theta \rangle
= (\scrI \eta \Psi | \xi^* \Theta \rangle
= (\scrI \eta \Phi | \xi^* \Theta \rangle
= (\eta \Phi | \xi^* \Theta \rangle
= (\eta \Phi | \Theta \rangle
\end{displaymath}
for all $\Theta \in \Cyl$, whence
$\eta \Psi = \eta \Phi$.
\end{quote}

\textit{Proof that $\scrI|_{\Cyl_{\diff}^*}$ is isometric and hence unitary:}
\begin{quote}
Let $\eta \Psi, \eta \Phi \in \Cyl_{\diff}^*$ be
given, so that $\Psi, \Phi \in \Cyl$. Using lemma \ref{cylorbit},
there exists $\varphi$ and $\xi$ in $\Diff$ such that
$\varphi^* \Psi, \xi^* \Phi \in \underline{\Cyl}$.  Using
the $\Diff$ invariance of $\eta$ and part (2.) of lemma \ref{avemaps_eq},
we have
\begin{eqnarray*}
\langle \scrI \eta \Psi,\scrI \eta \Phi \rangle
&=& \langle \scrI \eta (\varphi^* \Psi), \scrI \eta (\xi^* \Phi)
\rangle
= \langle \underline{\eta} (\varphi^* \Psi), \underline{\eta}(\xi^* \Phi) \rangle \\
&:=& (\underline{\eta}(\varphi^* \Psi)| \xi^* \Phi \rangle
= (\scrI \eta (\varphi^* \Psi)| \xi^* \Phi \rangle
= (\eta(\varphi^* \Psi) | \xi^* \Phi \rangle \\
&=& (\eta \Psi | \Phi \rangle
= \langle \eta \Psi, \eta \Phi \rangle,
\end{eqnarray*}
%
%
%
%
\end{quote}
\finishproof}
The above theorem implies
\begin{corollary}
$\Hil_{\diff}$ and $\underline{\Hil}_{\diff}$ are
isomorphic as Hilbert spaces.
\end{corollary}
\noindent It is then easy to extend the equivalence to the solution spaces solving both the
diffeomorphism and Gauss constraints:
\begin{corollary}
$\scrI|_{\Cyl_{\diff,G}^*}: \underline{\Cyl}_{\diff,G}^* \rightarrow \Cyl_{\diff,G}^*$
is a unitary isomorphism, so that $\Hil_{\diff,G}$ and $\underline{\Hil}_{\diff,G}$
are isomorphic as Hilbert spaces.
\end{corollary}
{\startproof
From the injectivity of $\scrI|_{\Cyl_\diff^*}$, we know $\scrI|_{\Cyl_{\diff,G}^*}$ is
injective.  It thus remains only to prove that $\scrI$ maps $\Cyl_{\diff,G}^*$ onto
$\underline{\Cyl}_{\diff,G}^*$, i.e., $\scrI[\Cyl_{\diff,G}^*] = \underline{\Cyl}_{\diff,G}^*$.\\
$(\subseteq)$:\\
Let $\eta \Psi \in \Cyl_{\diff, G}^*$ be given, so that $\Psi \in \Cyl \cap \Hil_G$, and
in particular $\Psi \in \Cyl_\gamma$ for some $\gamma \in \Gamma$.
By part (1.) of lemma \ref{cylorbit}, there exists $\varphi \in \Diff$ such that $\varphi \cdot \gamma \in \underline{\Gamma}$.
Then $(\varphi^{-1})^* \Psi \in \underline{\Cyl} \cap \underline{\Hil}_G$, and we have
$\scrI \eta \Psi = \scrI \eta (\varphi^{-1})^* \Psi = \underline{\eta} (\varphi^{-1})^* \Psi$,
where lemma \ref{eta_eq} was used in the second step. Thus
$\scrI \eta \Psi \in \underline{\Cyl}_{\diff, G}^*$.\\
$(\supseteq)$:\\
Let $\underline{\eta} \Psi \in \underline{\Cyl}_{\diff, G}^*$ be given, so that
$\Psi \in \underline{\Cyl} \cap \underline{\Hil}_G \subset \Cyl \cap \Hil_G$.  By lemma \ref{eta_eq},
$\scrI \eta \Psi = \underline{\eta} \Psi$, so that $\underline{\eta} \Psi \in \scrI[\Cyl_{\diff, G}^*]$.
\finishproof}

\section{Equivalence of diffeomorphism invariant operators, and equivalence
of dynamics}
\label{dynamsect}

When constructing operators in plLQG, we propose one quantize in exactly
the same way as in standard LQG, except that only piecewise linear edges
should be used.
For operators preserving $\Cyl$, this general statement can be made precise as follows.
Given an operator $\hat{O}_\omega$ in standard LQG,
a corresponding operator is defined in plLQG iff $\hat{O}_\omega$ preserves
$\underline{\Cyl}$, and in this case one defines the corresponding operator
$\hat{O}_{pl}$ in plLQG to be $\hat{O}_{\omega}|_{\underline{\Cyl}}$.  An
immediate consequence of this definition is
\begin{equation}
\label{scrI_intertwine}
\hat{O}_{pl}^* \circ \scrI = \scrI \circ \hat{O}_\omega^* .
\end{equation}
Next, we call an operator ``diffeomorphism invariant'' if it is invariant
under the group of generalized diffeomorphisms in the relevant framework.
If $\hat{O}_\omega$ preserves $\Cyl$ and is diffeomorphism invariant,
then it must be graph preserving\footnote{This can be seen as follows.
%
%
Suppose $\hat{O}$ is $\Diff$-invariant and preserves $\Cyl$.
Let $\Psi \in \Cyl_\alpha$ be given for some $\alpha$.  As $\hat{O}$ preserves $\Cyl$,
$\hat{O}\Psi \in \Cyl_\beta$ for some $\beta$.  From $\Diff$-invariance, we have
that for all $\varphi \in \Diff_\alpha$ (recall $\Diff_\alpha$ is the subgroup of  $\Diff$ preserving $\alpha$), $\hat{O}\Psi = U_\varphi \hat{O} U_{\varphi^{-1}} \Psi = U_\varphi \hat{O} \Psi$,
so that $\hat{O}\Psi \in \Cyl_{\varphi \cdot \beta}$ for all $\varphi \in \Diff_\alpha$.
Thus $\hat{O}\Psi \in \cup_{\varphi \in \Diff_\alpha} \Cyl_{\varphi \cdot \beta}$.  But given any
$\gamma, \gamma' \in \Gamma$,
%
%
$\Cyl_{\gamma} \cap \Cyl_{\gamma'} = \Cyl_{\gamma \cap \gamma'}$,
so that $\hat{O}\Psi \in \Cyl_{\cap_{\varphi \in \Diff_{\alpha}}\varphi \cdot \beta}$.
The only edges of $\beta$ that survive in $\cap_{\varphi \in \Diff_{\alpha}} \varphi \cdot \beta$
are those that are also edges of $\alpha$, whence in fact $\hat{O}\Psi \in \Cyl_\alpha$, showing
$\hat{O}$ is graph preserving.}
and hence also preserve $\underline{\Cyl}$,
so that there is a corresponding piecewise linear operator $\hat{O}_{pl}$.
Because $\underline{\Diff} \subset \Diff$, the $\Diff$-invariance of $\hat{O}_\omega$
also implies the $\underline{\Diff}$-invariance of $\hat{O}_{pl}$, so that
$\hat{O}_{pl}$ is diffeomorphism invariant.  These observations, along with
(\ref{scrI_intertwine}) allow us to state the following
\begin{proposition}
Given any diffeomorphism-invariant operator $\hat{O}_\omega$ preserving $\Cyl$
 in standard LQG, then $\hat{O}_\omega$ also preserves $\underline{\Cyl}$.
 The corresponding piecewise linear operator $\hat{O}_{pl}$ is also diffeomorphism invariant, and $\hat{O}_{pl}$ and $\hat{O}_\omega$ are mapped into
each other by the isomorphism $\scrI|_{\Cyl_{\diff}^*}$, that is,
\begin{equation}
\hat{O}_{pl}^* \circ \scrI = \scrI \circ \hat{O}_\omega^* .
\end{equation}
\end{proposition}
Note that, though this proposition seems quite general, in fact assuming $\hat{O}_\omega$
is both well defined on $\Cyl$ and diffeomorphism invariant is a relatively restrictive
assumption: it already constrains the applicability of the result to graph preserving operators. The master constraint operator \cite{master}, for example, though diffeomorphism invariant,
is not graph preserving.  This is possible because the master constraint is not well-defined on $\Cyl$, but rather must
be directly defined on $\Cyl_{\diff}^*$. We will later discuss the master constraint,
after we have discussed the Hamiltonian constraint.

The Hamiltonian constraint \cite{hamiltonian} is rather unique because it has as its domain
$\Cyl_{\diff}^*$, but does not map $\Cyl_{\diff}^*$ back into itself.
It is defined as follows. For each lapse $N$, each $\epsilon >0$ and each graph $\gamma$,
one defines a regulated operator $\hat{H}(N)_{\gamma, \epsilon}$ on $\Hort{\gamma}$.
Piecing these together for all $\gamma$ gives, for each $\epsilon$, an operator
$\hat{H}(N)_\epsilon$ on the kinematical Hilbert space $\Hil$.  The dual
$\hat{H}(N)_\epsilon^*$ then acts on $\Cyl^*$.  For any $\xi$ in
$\Cyl_\diff^* \subset \Cyl^*$, the limit
$\lim_{\epsilon \rightarrow 0} \hat{H}(N)_\epsilon^* \xi$
becomes trivial, allowing us to define
\begin{equation}
\hat{H}(N) \xi := \lim_{\epsilon \rightarrow 0} \hat{H}(N)_\epsilon^* \xi .
\end{equation}
$\hat{H}(N)$ is thus well-defined on $\Cyl_\diff^*$.
It is also diffeomorphism covariant:
$\left(U_\varphi^{-1}\right)^* \circ \hat{H}(N) \circ U_\varphi^*
= \hat{H}(\varphi^* N)$ for all $\varphi \in \Diff$, where $U_\varphi$ denotes
the unitary action of $\varphi$ on $\Hil$ via pullback.
However, for general lapse $N$, $\hat{H}(N)$ maps $\Cyl_\diff^*$ out of
itself due to $\hat{H}(N)$ not being diffeomorphism \textit{invariant}.
One can nevertheless define the solution to the Hamiltonian constraint to be
simply the common kernel of the operators $\hat{H}(N)$ for all lapse $N$.

This construction can be repeated in the obvious way for plLQG: one need only
ensure
that the loops added by the regulated $\hat{H}(N)_{\gamma, \epsilon}$ are
chosen
to be piecewise linear.  We do this, and then for $\gamma \in \underline{\Gamma}$, define
$\hat{\underline{H}}(N)_{\gamma,\epsilon}
:=\hat{H}(N)_{\gamma,\epsilon}|_{\underline{\Cyl}}$.
A construction exactly parallel to that above then goes through,
giving us a family of operators $\hat{\underline{H}}(N)$,
defined on $\underline{\Cyl}_\diff^*$, and diffeomorphism
covariant with respect to $\underline{\Diff}$, which nevertheless
generically map $\underline{\Cyl}_\diff^*$ out of itself.

Let $\ker \hat{H}$ denote the common kernel of the operators $\hat{H}(N)$
for all $N$,
and let $\ker \hat{\underline{H}}$ denote the common kernel of the operators
$\hat{\underline{H}}(N)$ for all $N$.
We have the following result:
\begin{proposition}
$\scrI|_{\ker \hat{H}}$ provides a unitary isomorphism
from $\ker \hat{H}$
onto $\ker \hat{\underline{H}}$.
\end{proposition}
{\startproof

We first note that for $\Psi \in \Cyl_\diff^*$, $\Phi \in \underline{\Cyl}$,
and any lapse $N$, the following relation holds:
\begin{eqnarray}
\nonumber
(\hat{\underline{H}}(N) \scrI \Psi | \Phi \rangle
&:=& \lim_{\epsilon \rightarrow 0}
(\scrI \Psi | \hat{\underline{H}}(N)_{\epsilon} \Phi \rangle \\
\nonumber
&=& \lim_{\epsilon \rightarrow 0} (\Psi | \hat{\underline{H}}(N)_{\epsilon} \Phi \rangle
= \lim_{\epsilon \rightarrow 0} (\Psi | \hat{H}(N)_{\epsilon} \Phi \rangle \\
\label{hamrel}
&=& (\hat{H}(N) \Psi | \Phi \rangle.
\end{eqnarray}
From this we immediately see that if
$\Psi \in \ker \hat{H}$, so that $\hat{H}(N)\Psi=0$ for all $N$,
then $\hat{\underline{H}}(N)\scrI\Psi=0$ for all $N$,
so that $\scrI \Psi \in \ker \hat{\underline{H}}$, whence
$\scrI[\ker \hat{H}] \subset \ker \hat{\underline{H}}$.

To prove the converse,
let $\Theta \in \ker \hat{\underline{H}}$ be given.  As $\underline{\Cyl}_\diff^*$
is defined to be the domain of the $\hat{\underline{H}}(N)$,
$\Theta \in \underline{\Cyl}_\diff^*$; using
the onto-ness of $\scrI|_{\Cyl_{\diff}^*}:\Cyl_\diff^* \rightarrow
\underline{\Cyl}_\diff^*$, there exists $\Psi \in \Cyl_\diff^*$ such that
$\Theta = \scrI \Psi$.  Next, let $N$ be given, and let $\Phi \in \Cyl$ be
given. By lemma \ref{cylorbit},
there exists $\varphi \in \Diff$ such that $\varphi^* \Phi \in \underline{\Cyl}$.
Using the $\Diff$ covariance of $\hat{H}(N)$ and then the
$\Diff$ invariance of $(\Psi|$,
\begin{eqnarray}
\nonumber
(\hat{H}(N)\Psi|\Phi\rangle
&=& ((U_\varphi^{-1})^* \circ \hat{H}((\varphi^{-1})^*N) \circ (U_\varphi)^*
\Psi|\Phi\rangle
= (\hat{H}((\varphi^{-1})^*N)\Psi| U_\varphi^{-1} |\Phi \rangle\\
&=& (\hat{H}((\varphi^{-1})^*N)\Psi|\varphi^* \Phi \rangle.
\end{eqnarray}
Applying relation (\ref{hamrel}) to $\varphi^* \Phi$
and $(\varphi^{-1})^* N$, and then using the fact that $\Theta = \scrI \Psi$ is
in $\ker \hat{H}$, the last line above is seen to be zero.  Thus
$(\hat{H}(N)\Psi|\Phi\rangle = 0$
for all $\Phi \in \Cyl$ and all lapse $N$, proving $\Psi \in \ker \hat{H}$, so that
$\Theta \in \scrI[\ker \hat{H}]$.  This proves the containment
$\ker \hat{\underline{H}} \subset \scrI[\ker \hat{H}]$, completing the
proof that $\ker \hat{\underline{H}} = \scrI[\ker \hat{H}]$.

As already shown in theorem \ref{diffiso},
$\scrI$ is injective and unitary on $\Cyl_\diff^*$, so that
it is also injective and unitary on $\ker \hat{H}$.  Thus
$\scrI|_{\ker \hat{H}} : \ker \hat{H} \rightarrow \ker \hat{\underline{H}}$
provides a unitary isomorphism between $\ker \hat{H}$ and $\ker \hat{\underline{H}}$.
\finishproof}
Finally, the physical Hilbert space of solutions to the diffeomorphism, Gauss,
\textit{and} Hamiltonian constraint
in LQG and plLQG are $\Hil_{\Phys}:=\overline{\Cyl_{\diff,G}^*\cap \ker \hat{H}}$ and
$\underline{\Hil}_{\Phys}:= \overline{\underline{\Cyl}_{\diff,G}^* \cap \ker \hat{\underline{H}}}$, respectively, where the closure denotes Cauchy completion.
As the isomorphism $\scrI|_{\Cyl_{\diff}^*}$ maps the inner product on $\Cyl_{\diff}^*$ onto that on
$\underline{\Cyl}_{\diff}^*$, maps $\Cyl_{\diff,G}^*$ onto $\underline{\Cyl}_{\diff,G}^*$,
and maps $\ker \hat{H}$ onto $\ker \hat{\underline{H}}$, it is immediate that $\scrI$ provides
a unitary isomorphism between these physical Hilbert spaces.

We now come to the master constraint.  Let us review its construction in standard LQG
from \cite{master}.
%
%
%
First, given a spatial point $v \in M$, let $N_v(x):= \delta_{v,x}$,
a particular singular choice of lapse. The corresponding Hamiltonian constraint operator
$\hat{H}_v := \hat{H}(N_v)$ is nevertheless well defined \cite{hamiltonian},
as is perhaps not surprising given the discreteness of LQG. We next recall the \textit{generalized spin-network} functions $T_\sigma$, where $\sigma$ denotes the
triple $(\gamma, \vec{j}, \vec{T})$ of a graph $\gamma \in \Gamma$,
an assignment of a spin to each edge, and an assignment of a tensor among representations to each
vertex
\cite{alrev, otherlqg}.  We require that all spin labels be non-trivial.  Furthermore,
as in, e.g., \cite{alrev}, for each possible set of representations incident at a vertex,
we fixed a basis of the tensor space among the representations. Let $\scrS$ denote the space of all such triples $(\gamma, \vec{j}, \vec{T})$.
$\{T_\sigma\}_{\sigma \in \scrS}$ forms an orthonormal basis of $\Cyl$ and hence $\Hil$.
Furthermore, $\Diff$ acts on $\scrS$, so that we may consider the
$\Diff$-equivalence class of an element $\sigma \in \scrS$, which we denote $[\sigma]_{\Diff}$.
With these definitions made,
we define a quadratic form $Q_M: \Cyl_\diff^* \times \Cyl_\diff^* \rightarrow \C$
by
\begin{equation}
Q_M(\Phi, \Psi):= \sum_{[\sigma]_\Diff} \eta_{[\sigma]_\Diff}
\sum_{v \in V(\gamma(\sigma))} \overline{(\hat{H}_v \Phi | T_{\sigma}\rangle}
(\hat{H}_v \Psi | T_{\sigma}\rangle.
\end{equation}
where $\eta_{[\sigma]_\Diff}:= 1/|GS_{\gamma(\sigma)}|$ are the coefficients appearing
in the last expression in (\ref{an_eta}) for the diffeomorphism constraint rigging map,
and where $V(\gamma(\sigma))$ denotes the set of vertices in
$\gamma(\sigma)$.\footnote{In \cite{master},
$\eta_{[\sigma]_\Diff}$ are a set of constants parametrizing
an ambiguity in the definition of the rigging map discussed in the original
work \cite{almmt}. Here, as earlier in this paper, we are taking a natural resolution to
this ambiguity suggested in \cite{alrev}, leading to the specific values of
$\eta_{[\sigma]_\Diff}$ given above.}
$Q_M(\cdot, \cdot)$ then determines the master constraint $\hat{M}$ uniquely
via \cite{master}
\begin{equation}
\hat{M} \Phi := \sum_{x \in I} Q_M(B_x,\Phi) B_x
\end{equation}
where $\{B_x\}_{x\in I}$ is any orthonormal basis of $\Cyl_\diff^*$.

A construction parallel to the above goes through in the plLQG case.
Let $\underline{\scrS}$ denote the set of generalized spin-network labels
$\sigma=(\gamma, \vec{j}, \vec{T})$
such that $\gamma \in \underline{\Gamma}$.  Then $\underline{\Diff}$ acts on $\underline{\scrS}$,
so that for each $\sigma \in \underline{\Diff}$, one can define an equivalence class $[\sigma]_{\underline{\Diff}}$.
The quadratic form for the piecewise linear framework is then
\begin{equation}
Q_{\underline{M}}(\Phi, \Psi):= \sum_{[\sigma]_{\underline{\Diff}}} \eta_{[\sigma]_{\underline{\Diff}}}
\sum_{v \in V(\gamma(\sigma))} \overline{(\hat{\underline{H}}_v \Phi | T_{\sigma}\rangle}
(\hat{\underline{H}}_v \Psi | T_{\sigma}\rangle
\end{equation}
where $\Phi, \Psi \in \underline{\Cyl}_\diff^*$, and where
$\eta_{[\sigma]_{\underline{\Diff}}} = 1/|\underline{GS}_{\gamma(\sigma)}|$ are the coefficients
in the plLQG rigging map (\ref{pl_eta}). The master constraint is then
\begin{equation}
\hat{M} \Phi := \sum_{x \in I} Q_M(\underline{B}_x,\Phi) \underline{B}_x
\end{equation}
where $\{\underline{B}_x\}_{x\in I}$ is any orthonormal basis of $\underline{\Cyl}_\diff^*$.

\begin{proposition}
$\hat{M}$ is mapped into $\hat{\underline{M}}$ by the isomorphism $\scrI|_{\Cyl_\diff^*}$.
\end{proposition}
{\startproof
In each case the master constraint is determined from the
quadratic form and inner product on diffeomorphism invariant states
in the same way.  To prove equivalence of the master constraints,
%
%
it is thus sufficient to prove equivalence of the quadratic forms; that is,
we want to show $Q_{\underline{M}}(\scrI \Phi, \scrI \Psi) = Q_M(\Phi, \Psi)$
for all $\Phi, \Psi \in \Cyl_\diff^*$:
\begin{eqnarray}
\nonumber
Q_{\underline{M}}(\scrI \Phi, \scrI \Psi)
&:=& \sum_{[\sigma]_{\underline{\Diff}}} \eta_{[\sigma]_{\underline{\Diff}}}
\sum_{v \in V(\gamma(\sigma))} \overline{(\hat{\underline{H}}_v \scrI \Phi | T_{\sigma}\rangle}
(\hat{\underline{H}}_v \scrI \Psi | T_{\sigma}\rangle \\
\nonumber
&=& \sum_{[\sigma]_{\underline{\Diff}}} \eta_{[\sigma]_{\underline{\Diff}}}
\sum_{v \in V(\gamma(\sigma))} \lim_{\epsilon, \epsilon' \rightarrow \infty}
\overline{(\scrI \Phi | \hat{\underline{H}}_{v,\epsilon} T_{\sigma}\rangle}
(\scrI \Psi | \hat{\underline{H}}_{v,\epsilon} T_{\sigma}\rangle \\
\nonumber
&=& \sum_{[\sigma]_{\underline{\Diff}}} \eta_{[\sigma]_{\underline{\Diff}}}
\sum_{v \in V(\gamma(\sigma))} \lim_{\epsilon, \epsilon' \rightarrow \infty}
\overline{(\Phi | \hat{H}_{v,\epsilon} T_{\sigma}\rangle}
(\Psi | \hat{H}_{v,\epsilon} T_{\sigma}\rangle \\
\label{Qderiv}
&=&\sum_{[\sigma]_{\underline{\Diff}}} \eta_{[\sigma]_{\underline{\Diff}}}
\sum_{v \in V(\gamma(\sigma))} \overline{(\hat{H}_v \Phi | T_{\sigma}\rangle}
(\hat{H}_v \Psi | T_{\sigma}\rangle
\end{eqnarray}
where, in the third equality, we have used the definition of $\scrI$ and that
$\hat{\underline{H}}_{v,\epsilon} = \hat{H}_{v,\epsilon}|_{\underline{\Cyl}}$.
Now, the outer sum in (\ref{Qderiv}) is over $[\sigma]_{\underline{\Diff}} \in \underline{\scrS}/\underline{\Diff}$.
Define $J: \underline{\scrS}/\underline{\Diff} \rightarrow \scrS/\Diff$ by
$[\sigma]_{\underline{\Diff}} \mapsto [\sigma]_{\Diff}$.  $J$ is well-defined due to
$\underline{\Diff} \subset \Diff$.  Using lemma \ref{plgraphtrans}, one shows that it is 1-1,
and using lemma \ref{cylorbit} one sees that it is onto. (Details: exercise for the reader.)
Furthermore, as $\sigma\in\underline{\scrS}$, $\gamma(\sigma) \in \underline{\Gamma}$,
so that from lemma \ref{Fiso}, $|\underline{GS}_{\gamma(\sigma)}| = |GS_{\gamma(\sigma)}|$,
and we have $\eta_{[\sigma]_{\underline{\Diff}}} = \eta_{[\sigma]_{\Diff}}$.  Using the
isomorphism $J$ to replace
$[\sigma]_{\underline{\Diff}}$ with $[\sigma]_{\Diff}$ in (\ref{Qderiv}),
we obtain
\begin{equation}
Q_{\underline{M}}(\scrI \Phi, \scrI \Psi) = Q_{M}(\Phi, \Psi).
\end{equation}
\finishproof}
Lastly, because $\scrI|_{\Cyl_\diff^*}$ maps the master constraint $\hat{M}$ onto
$\hat{\underline{M}}$, and $\Cyl_{\diff,G}^*$ onto $\underline{\Cyl}_{\diff,G}^*$, $\scrI|_{\Cyl_{\diff,G}^*}$ will map $\hat{M}|_{\Cyl_{\diff, G}^*}$ onto $\hat{\underline{M}}|_{\underline{\Cyl}_{\diff, G}^*}$, so that the master constraint
dynamics are also equivalent after solving both the diffeomorphism and Gauss
constraints.\footnote{In \cite{master}, the master constraint is in fact constructed
directly on $\Cyl_{\diff,G}^*$.}

The above results show that not only are the diffeomorphism invariant Hilbert spaces in
LQG and plLQG unitarily isomorphic, but the dynamics (whether defined with Master or Hamiltonian constraint) are isomorphic as well, so that the two frameworks are truly equivalent.

\section{Exact embedding of LQC into piecewise linear LQG}
\label{embed_sect}

In the paper \cite{engle2007}, an embedding of LQC into
the space $\underline{\Cyl}^*$ was constructed.
In that context,
the space $\underline{\Cyl}^*$ was unnatural as a distributional
space in the sense that it was the dual of a test function space that
is not dense in the traditional kinematical Hilbert space $\Hil$ of LQG.
A possible physical meaning for $\underline{\Cyl}^*$ was suggested
in \cite{engle2007}, but this did not solve the fact that it was not
clear how to use $\underline{\Cyl}^*$ for the next step in the program
of \cite{engle2006, engle2007}.  Specifically, the next step was to
group average the kinematical embeddings to obtain embeddings into
LQG at the diffeomorphism invariant level. Note
one \textit{must} construct embeddings into LQG at the diffeomorphism invariant level
if one hopes to exactly relate the Hamiltonian constraints in LQC and LQG in any way,
as the latter is defined only on diffeomorphism invariant states.
To accomplish the construction of the diffeomorphism invariant embeddings, two
issues needed to be addressed \cite{engle2007}:
\begin{enumerate}
\item The group of piecewise analytic diffeomorphisms did not
even act on $\underline{\Cyl}^*$, so that one could not even write
down a formal expression for group averaging the kinematical embeddings.

\item Once one is able to write down a formal group averaging, one
would need to regulate the integral over diffeomorphisms in some way.
\end{enumerate}
It is in this first step that the use of $\underline{\Cyl}^*$ seemed to prevent
further progress.

In the construction of plLQG, $\underline{\Cyl}^*$ also appears, but this time
\textit{as the space of distributional states for a completely
parallel framework for loop quantum gravity, which, as was proven
above, is equivalent to the standard one at the diffeomorphism-invariant
level.} Furthermore, the space of `piecewise linear generalized diffeomorphisms'
acts on $\underline{\Cyl}^*$, so that one can now formally write down
the group averaging of the embeddings, providing an expression
for the embedding into the space of diffeomorphism invariant states.
Because of the isomorphism between plLQG and LQG at the diffeomorphism
invariant level, this is also a formal expression for the embedding into the space
of diffeomorphism invariant states in standard LQG.
That is, the first obstruction listed above is gone.
Because the embeddings of \cite{engle2007} were a motivation for the
present work, we briefly review them here; we then end the section with the new
expressions for the diffeomorphism invariant embeddings.

First we recall some necessary structures from loop quantum cosmology (LQG).
As in \cite{abl}, we take the classical configuration space for homogeneous,
isotropic cosmology to be the space of homogeneous, isotropic connections, but in
the gauge-fixed sense defined in \cite{engle2007}; we denote this space by $\scrA_S$.
By picking a reference connection $\mathring{A}^i_a \in \scrA_S$,
all other connections in $\scrA_S$ are related to $\mathring{A}^i_a$ by scaling.
Thus if we define $r: \R \rightarrow \scrA_S \subset \scrA$ by
\begin{equation}
r: c \mapsto c \mathring{A}^i_a,
\end{equation}
$r$ provides an isomorphism of $\R$ with $\scrA_S$.
States in LQC are then functions on $\R \cong \scrA_S$.
The basic space of `nice' states in LQC (and one of the sources of
the unique character of LQC) is the space of
\textit{almost periodic functions}; following \cite{abl}, we denote this $\Cyl_S$.
$\Cyl_S^*$ is the space of distributional states.

The kinematical and gauge-invariant embeddings of \cite{engle2007} are
then defined as follows.  The `c' embedding $\iota_c: \Cyl_S^* \rightarrow \underline{\Cyl}^*$
is defined by
\begin{equation}
(\iota_c \psi| \Phi\rangle := (\psi | r^* \Phi \rangle .
\end{equation}
From $\iota_c$, one constructs the `b' embeddings.
To remind the reader from \cite{engle2006, engle2007}, the `b'
embedding is built using coherent states, the idea being to
use the freedom in the choice of coherent states to adapt the
embedding to be approximately preserved by the dynamics.
In \cite{engle2006, engle2007}, complexifier
coherent states are used; in complexifier coherent states,
the freedom in choosing the family of coherent states
is parametrized by a choice of complexifier \cite{complexifier}.  To introduce the
complexifiers, first let $X_S$ and $X$
denote the classical phase space of the reduced and full theories, respectively.
Then let $C_S: X_S \rightarrow \R^+$, $C: X \rightarrow \R^+$ be
any two (pure momentum) complexifiers \cite{complexifier}.
%
%
Let $\hat{C}_S$ and $\hat{C}$ denote their respective quantizations in the reduced and full
quantum theories.  For brevity, we give only the final expression for
the corresponding `b' embedding
$\iota_b: \Cyl_S^* \rightarrow \underline{\Cyl}^*$.  It is given by \cite{engle2007}
\begin{equation}
(\iota_b \psi |\Phi\rangle
:= (\psi| e^{\hat{C}_S} \circ r^* \circ e^{-\hat{C}} | \Phi \rangle.
\end{equation}
The Gauss-gauge invariant versions of these embeddings are
$\iota_c^G:= P_G^* \circ \iota_c$ and $\iota_b^G:= P_G^* \circ \iota_b$,
where $P_G: \underline{\Cyl} \rightarrow \underline{\Cyl}$ denotes
the projector onto gauge-invariant states.
For the motivation behind these definitions and their nice properties,
we refer the reader to the original papers \cite{engle2006, engle2007}.

Now we come to the formal expression for the embedding into diffeomorphism
invariant states, made possible by the new piecewise LQG framework introduced in this
paper. The diffeomorphism invariant embedding $\iota_c^{Diff}: \Cyl_S^* \rightarrow \underline{\Cyl}_{\diff}^*$ has the formal expression
\begin{eqnarray}
\nonumber
(\iota_c^{Diff} \psi |\Phi\rangle
&:=& \left(\int_{\varphi \in \underline{\Diff}} \scrD \varphi \,
(U_{\varphi}^* \iota_c^G \psi|\right)|\Phi\rangle \\
\nonumber
&=& \int_{\varphi \in \underline{\Diff}} \scrD \varphi \, (\iota_c^G \psi|
U_\varphi |\Phi \rangle \\
\label{diffc}
&=& \int_{\varphi \in \underline{\Diff}} \scrD \varphi \, (\psi| r^* \circ P_G \circ U_{\varphi} |\Phi \rangle .
\end{eqnarray}
The formal expression for the diffeomorphism invariant `b' embedding
$\iota_b^{Diff}: \Cyl_S^* \rightarrow \underline{\Cyl}_{\diff}^*$ is then
\begin{equation}
\label{diffb}
(\iota_b^{Diff} \psi |\Phi\rangle
= \int_{\varphi \in \underline{\Diff}} \scrD \varphi \, (\psi| e^{\hat{C}_S} \circ r^* \circ e^{-\hat{C}}
\circ P_G \circ U_{\varphi} |\Phi \rangle .
\end{equation}
If $\hat{C}$ and $\hat{C}_S$  are gauge and diffeomorphism invariant,
this reduces to $\iota_b^{Diff} = e^{-\hat{C}^*} \circ \iota_c^{Diff}
\circ e^{\hat{C}_S^*}$.
Composing (\ref{diffc}) and (\ref{diffb}) with the isomorphism $\underline{\Cyl}_{\diff}^* \leftrightarrow \Cyl_{\diff}^*$ defined in section \ref{equivsect} then provides us with the formal expression for the `c' embedding into $\Cyl_{\diff}^*$, and for the `b' embeddings into $\Cyl_{\diff}^*$.
The use of $\underline{\Diff}$ instead of $\Diff$ not only has allowed us to write these
expressions, but the fact that $\underline{\Diff}$ is so much smaller than $\Diff$ makes
it more likely that they can be regularized.

\section{Discussion}

The kinematics of LQG are usually formulated in terms of the piecewise analytic
category. We have shown that the piecewise analytic category is not essential, and
can be replaced with something as simple as the piecewise linear category,
giving rise to what we have called \textit{piecewise linear LQG} (plLQG).
We have shown that piecewise linear LQG is fully equivalent to standard LQG
at the diffeomorphism invariant level, both in terms of Hilbert space structure
and dynamics, as long as one makes a natural choice of
generalized diffeomorphism group such as advocated in \cite{koslowski2006}.

Furthermore, we have seen that LQC is exactly embeddable into plLQG.
This shows that the non-embeddability result of \cite{bf2007}
is perhaps somewhat of a red herring: it appears relevant at the kinematical level,
but this relevance seems to evaporate at the diffeomorphism invariant level.
For, plLQG circumvents
the non-embeddability result of \cite{bf2007}, and is yet fully equivalent to LQG at
the diffeomorphism invariant level. This is what has now allowed us
to at least write down formal expressions for embeddings of LQC into LQG at the
diffeomorphism invariant level.  These expressions were given in section
\ref{embed_sect}.\footnote{As
a side note, it may also be possible that there is another way to relate LQG to cosmology
other than via the piecewise linear framework presented here. For, as pointed out by
Koslowski \cite{koslowski_priv}, it appears that, given any analytic edge $e$, the
holonomy along $e$ as a function of the symmetric connection $A_S = r(c) = c \mathring{A}^i_a$
can be decomposed into an almost periodic part \cite{abl}
and a part vanishing as $c$ approaches infinity.
If true, it is not hard to see that this decomposition must be unique,
as there are no almost periodic functions that vanish at infinity. This would then allow
one to construct a projector $P_{\rm ap}: r^*[\Cyl] \rightarrow \Cyl_S$
that projects out the almost periodic part. The projector could then be used
to construct embeddings $\iota_c$ and $\iota_b$ of LQC directly into $\Cyl^*$:
$(\iota_c \psi | \Phi\rangle := (\psi | P_{\rm ap} r^* | \Phi \rangle$, and then
$\iota_b:= e^{-\hat{C}^*} \circ \iota_c \circ e^{\hat{C}_S^*}$.  These
embeddings would again satisfy the physical intertwining criterion used in
\cite{engle2007}. As the codomain of such embeddings would be directly $\Cyl^*$,
and $\Diff$ acts on $\Cyl^*$, one would then be able to directly write down a
formal expression for `c' and `b' embeddings into diffeomorphism invariant
states, similar to that in section \ref{embed_sect} of this paper. One could
then check whether the resulting formal embedding is equivalent to the one given in this paper.
Of course, the resulting embedding would also have to be regularized.
For the present, this is just a future possibility.}
Of course it still remains to regulate these expressions in some way.

We close with some remarks regarding the similarities of piecewise linear
LQG to the framework underlying the contruction of spinfoams.  As argued, for example,
in \cite{epr}, the classical theory underlying spinfoams is a certain
discrete theory based on piecewise flat geometries.  Furthermore, as touched
upon in appendix B of \cite{epr}, in order for the discrete variables
to fully describe the piecewise flat geometry, one is implicitly assuming
a given linear structure on each patch. Thus, one is actually assuming a
piecewise linear structure of spacetime.  As seen in this paper, the use
of piecewise linear structures naturally leads to the use of simplicial complexes,
and simplicial complexes are central in the classical discrete theory underlying
spinfoams.  Whether the relation between plLQG and spinfoams
goes beyond these cursory remarks is not clear, and would be interesting
to investigate.

\section*{Acknowledgements}
The author thanks Tim Koslowski for stimulating
exchanges, Andrew Ranicki for encouraging him to study simplicial complexes
in more depth, and Thomas Thiemann for remarks on a prior draft.
This work was supported in part by the Alexander von Humboldt
foundation of Germany.

\appendix

\section{Existence of triangulation compatible with a graph}

We prove here a result that is needed in section \ref{equivsect} for proving
the unitary isomorphism between the diffeomorphism invariant Hilbert spaces of
plLQG and LQG.
%
%
We place it in an appendix because it requires
a number of new definitions that are not needed elsewhere in the paper, and would distract
from the logic of section \ref{equivsect}.

For the purposes of this appendix, we remind the reader, from
section \ref{equivsect}, that a 1-complex $X$ is said to be \textit{compatible} with a piecewise linear
graph $\gamma$ if $|X|$ is equal to the image of $\gamma$.
In the following, we will also need the notion of a \textit{cell} and a \textit{cell complex},
which we define here.  To summarize in short, a \textit{cell} is a compact convex polyhedron (see \cite{rs} for a definition in terms of more rudimentary notions).  Given a cell $C$, and a plane $P$ such that
$C \setminus P$ is connected, we call $A:=P \cap C$ a \textit{face} of $C$,
and we write $A<C$.
The vertices, edges, faces of $C$ in the usual sense, as well as $C$ itself, are
are all faces of $C$. A \textit{cell complex} is defined in a manner analogous to
a simplicial complex:
\begin{definition}[cell complex]
A \textit{cell complex} $K$ is a finite collection of cells satisfying
(i) if $C \in K$ and $B$ is a face of $C$, then $B \in K$ and (ii) If $B,C \in K$,
then $B \cap C$ is a face of $B$ and $C$.
\end{definition}
\noindent Given a cell complex $K$, we define $|K|:=\cup_{A \in K} A$
as the polyhedron
underlying $K$.
Given two points $p$,$q$, let $pq$ denote the line segment between them.
Then given a cell $A$ and a point $p$ not in the plane determined
by $A$, one defines the \textit{cone} with vertex $p$ and base $A$,
denoted $pA$, by $pA:=\cup_{q \in A} (pq)$.  Given two cell complexes $K$ and $L$,
$K$ is said to be a \textit{subdivision}
of $L$ if $|K|=|L|$ and every cell in $K$ is contained in a cell of $L$.
A subdivision $K$ of $L$ is said to be obtained by
\textit{starring} at a point $a$ if $K$ is obtained from $L$ by replacing each cell
$C \in L$ with $a \in C$ by the collection of cells $\{aF | F < C, a \notin F\}$
(see p.15 of \cite{rs}).

With these preliminaries out of the way, we come to the theorem.
\begin{theorem}
\label{triangexists}
Given any piecewise linear graph $\gamma$, there exists a triangulation $K$
of $\R^3$ containing a 1-dimensional subcomplex $K_1$ compatible with $\gamma$.
\end{theorem}
{\startproof
Let $X$ be the minimal 1-complex compatible with $\gamma$: that is, break up
each edge of $\gamma$ into its straight parts, and then define $X$ to contain all
of these straight parts and their end points.  Let $N$ denote any rectangular prism
sufficiently large so that it contains all of $X$, without $X$ intersecting the boundary of
$N$.

For each 1-simplex $e$ in $X$, let $e_1, e_2$ denote the end points.
%
%
Construct a cell complex $N_{e_1}$ by starring $N$ at the point $e_1$.
Then construct $N_e$ by starring $N_{e_1}$ at $e_2$.
Because every cell of $N_{e_1}$ containing $e_2$ possesses $e_1$
as a vertex, the starring procedure gaurantees
$e_2 e_1 = e$ will belong to $N_e$.  Take the repeated \textit{intersection}
of the cell complexes $N_e$,
\begin{equation}
Q:= \{\cap_{e \in X} A_e | \{A_e \in N_e\}_{e \in X}\} .
\end{equation}
Noting example 2.8(5) of \cite{rs}, this is again a cell complex.
Furthermore, $Q$ is a subdivision of each cell complex $A_e$.  It therefore
contains a subdivision $\tilde{e}$ of each $e$.  Taking the union of these
subdivisions $\tilde{e}$ provides
a 1-complex $K_1$ that is a subcomplex of $Q$, and that is compatible with
$\gamma$.
Next, from proposition 2.9 of \cite{rs}, $Q$ can be subdivided further to
obtain a simplicial complex $H$, without adding any vertices, so that $K_1$ is again a subcomplex
of $H$.  Now, $|H| = N$ is a rectangular prism.  Choose a vertex $v$ of $N$, and let
$P_1, P_2, P_3$ denote the three planes passing through $v$ that contain two dimensional
faces of $N$.  By reflecting $H$ repeatedly across these three planes, we obtain seven further
copies of $H$ that match on their common boundaries among themselves and with $H$.  The union
of $H$ with these copies therefore defines a simplicial complex $J$.  $J$ is furthermore such that
if we introduce an infinite number of copies of it, tiling all of $\R^3$, these copies will match
on their common boundaries.  If we let $K$ denote the union of $J$ with this infinite number of copies
of $J$, then $K$ is a simplicial complex.  $K$ triangulates all of $\R^3$, and contains the
one dimensional subcomplex $K_1$, which is compatible with $\gamma$.
\finishproof}

\end{document}